\documentclass[12pt]{wlscirep}
\title{Transient chaos - a resolution of breakdown of quantum-classical 
correspondence in optomechanics}

\author[1]{Guanglei Wang}
\author[1,2,3,*]{Ying-Cheng Lai}
\author[3]{Celso Grebogi}
\affil[1]{School of Electrical, Computer, and Energy Engineering,
Arizona State University, Tempe, Arizona 85287, USA}
\affil[2]{Department of Physics, Arizona State University, Tempe,
Arizona 85287, USA}
\affil[3]{Institute for Complex Systems and Mathematical Biology,
King's College, University of Aberdeen, Aberdeen AB24 3UE, UK}
\affil[*]{Ying-Cheng.Lai@asu.edu}


\begin{abstract}

Recently, the phenomenon of {\em quantum-classical correspondence 
breakdown} was uncovered in optomechanics, where in the classical regime 
the system exhibits chaos but in the corresponding quantum regime 
the motion is regular - there appears to be no signature of
classical chaos whatsoever in the corresponding quantum system,
generating a paradox. We find that transient chaos, besides being a 
physically meaningful phenomenon by itself, provides a resolution. 
Using the method of quantum state diffusion to simulate the system 
dynamics subject to continuous homodyne detection, we uncover 
transient chaos associated with quantum trajectories. 
The transient behavior is consistent with chaos in the classical 
limit, while the long term evolution of the quantum system is regular. 
Transient chaos thus serves as a bridge for the quantum-classical 
transition (QCT). Strikingly, as the system transitions from the quantum to 
the classical regime, the average chaotic transient lifetime increases 
dramatically (faster than the Ehrenfest time characterizing the QCT  
for isolated quantum systems). We develop a physical theory to explain 
the scaling law.

\end{abstract}

\begin{document}

\flushbottom
\maketitle
\thispagestyle{empty}

\section*{Introduction} \label{sec:intro}

The quantum-classical correspondence is a fundamental and fascinating
problem in physics. For a specific physical process in a quantum system, 
if a large number of energy levels are involved (e.g., in the high 
energy regime), the evolution of the expected values of the observables 
will be governed by the classical Newtonian dynamics. This is the 
usual quantum-classical correspondence. Exceptions can 
occur when only a few lower energy levels are involved, e.g., at low 
temperatures, such that the quantum features of the ground state are 
manifested on a macroscopic scale~\cite{Feymann:book}, leading to 
fascinating phenomena such as Bose-Einstein condensation, 
superconductivity, and superfluids. In this paper, we report our discovery 
of transient chaos as a natural paradigm to explain the recently discovered 
phenomenon of the breakdown of quantum-classical correspondence in 
optomechanics.

\begin{figure}
\centering
\includegraphics[width=0.8\linewidth]{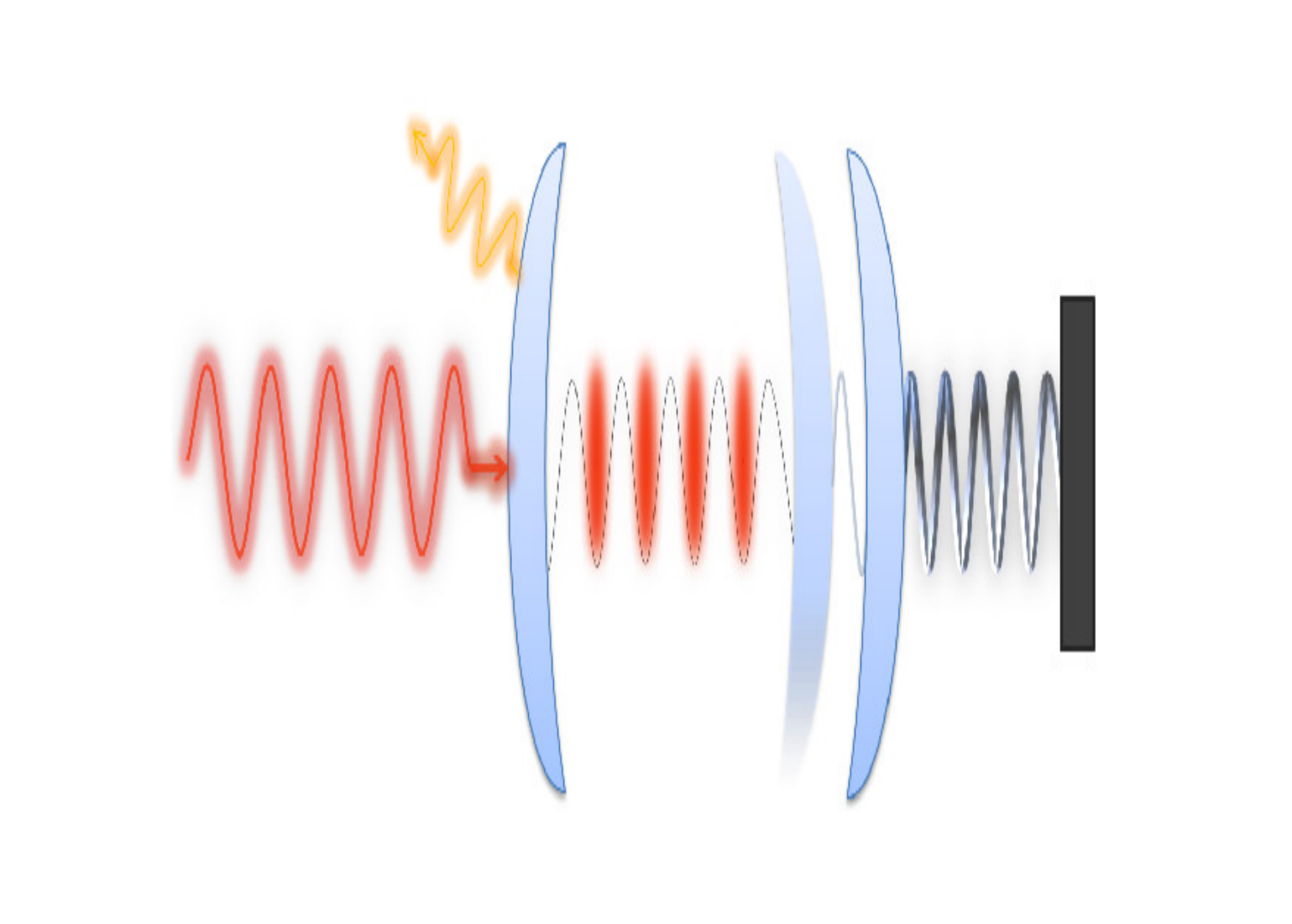}
\caption{A schematic figure of the optomechanical system.}
\label{fig:Sch}
\end{figure}
 
A prototypical optomechanical system consists of an optical cavity with 
a fixed mirror and a nanoscale, mechanically movable cantilever, as shown 
schematically in Fig.~\ref{fig:Sch}. The basic physics is that the radiation 
pressure from the optical field changes the position of the movable mirror, 
which in return modulates the resonance frequency of the optical cavity, 
leading to a coupling between the optical and mechanical degrees of 
freedom~\cite{MG:2009,AKM:2014}. In addition to this prototypical setting, 
alternative configurations for realizing the optical-mechanical coupling 
exist, such as those based on the whispering-gallery modes~\cite{ASY:2014}, 
microtoroid~\cite{GLPV:2010} and microsphere~\cite{AKM:2014} resonators. 
Optomechanics is thus not only fundamentally important, as it provides a 
setting to understand the physics of optical-mechanical 
interactions~\cite{LPXBHT:2008,LPT:2009,AKM:2014}, but also 
practically significant with applications ranging from ultra-precision 
measurements~\cite{MG:2009,JARAK:2009,VTBCH:2010}, light-matter 
entanglement~\cite{VGFBTGVZA:2007,GHVA:2009,ME:2009},
mechanical memory~\cite{BPLPT:2011}, tunable optical coupler~\cite{FPLT:2011}, 
classical state preparation through squeezing~\cite{PFT:2014,PFT:2015},
optical transparency~\cite{FFPT:2015}, and photon shuttling~\cite{LL:2014} to 
creation of nonclassical light~\cite{BBSPBS:2012,QCHM:2012} and cooling of
microscopic or mesoscopic objects~\cite{TDLHACSWLS:2011,CASHKGAP:2011}.
The classical equations of motion of an optomechanical system are 
nonlinear, rendering possible chaotic behaviors~\cite{CV:2007,CCV:2007}. 

In a recent work~\cite{BAF:2015}, it was demonstrated that, in the 
classical regime where the system exhibits chaos, in the corresponding 
quantum regime the motion becomes regular and no signatures of chaos 
appear to exist. This is the so-called {\em quantum-classical 
correspondence breakdown} in optomechanics.
A conventional approach to studying the correspondence is 
to compare the quantum Wigner function distribution with the classical 
phase space distribution~\cite{KRSL:2007,KLRS:2008,QCHM:2012}, both 
being average quantities. However, a recent work~\cite{WHHRHA:2015} 
demonstrated an optimal state estimation for cavity optomechanical 
systems through Kalman filtering, which allows us to obtain the conditional 
system state in the presence of experimental noise. In addition, 
observation of quantum trajectories obeying quantum state diffusion 
through heterodyne detection in a coupled system between a superconducting 
qubit and an off-resonant cavity was reported~\cite{CSBSMRH:2016}, as
well as other types of quantum trajectories~\cite{GKGBDHBRH:2007,
YZSWDCWH:2008,VLMZFBA:2010,NBSRFHWJ:2010,VSS:2011,MWMS:2013,
HSMSGBSAFG:2013,VPSAWBGSHC:2014,FMDS:2014,WCDJMS:2014,LRTETJWSD:2014}.
Thus, rather than focusing on the average properties of the system, we 
study the individual quantum trajectories of the system as related to the 
continuous weak measurement to probe into the quantum-classical 
correspondence breakdown.

Our aim is to uncover, through systematic classical and quantum 
simulations, the dynamical and physical mechanisms responsible for the 
breakdown phenomenon. The standard treatment~\cite{AKM:2014} of an 
optomechanical system consists of quantizing the cavity optical field 
and the oscillations of the cantilever as two mutually interacting 
quantum boson fields while treating the driving laser field classically. 
Dissipation associated with the optical and mechanical
fields can be incorporated into the quantum Langevin equations from the
quantum input-output theory~\cite{GZ:book} or by solving the quantum Master
equation with the Lindblad operators. When chaos occurs in the classical limit,
the system is typically in a high energy state with hundreds of photons 
and phonons, rendering infeasible direct simulation of the quantum 
Master equation. An effective framework is the method of quantum state
diffusion (QSD), which generates quantum trajectories to
approximate the time evolution as governed by the quantum Master
equation~\cite{GP:1992,GP:1993a,GP:1993b}. The QSD method has been
instrumental to homodyne detection and the study of quantum-classical 
correspondence in dissipative quantum chaos~\cite{BPS:1996,BHJS:2002,KP:2008}. 
Here, using the QSD method, we calculate the dynamical trajectories 
of the system in the quantum regime. Our computations extending to 
the long time scales (which were not attempted in previous works) 
suggest that transient chaos~\cite{LT:book} associated with quantum 
trajectories is ubiquitous. (To our knowledge, in spite of reports of
chaos~\cite{CV:2007,CCV:2007,AKM:2014}, there were no prior results of
transient chaos in optomechanical systems.) In particular, before 
approaching a regular final state, the quantum system exhibits a behavior 
that is consistent with the classical chaotic behavior. Thus, in short and 
in long time scales, the time evolutions of the system in the quantum regime 
would appear to be chaotic and regular, respectively. This means that, 
in short time scales a quantum-classical correspondence does exist, but 
its breakdown occurs in the long time limit. A striking finding is that, 
as the classical regime is approached, the average transient lifetime 
increases dramatically (faster than the Ehrenfest time - see Discussion). 
As the quantum system becomes ``more classical,'' the quantum-classical 
correspondence holds significantly longer, providing a natural 
resolution for the breakdown phenomenon. 

\section*{Results} \label{sec:results}

\paragraph*{Hamiltonian.}
In the rotating frame of the driving laser field, the Hamiltonian of a 
generic optomechanical system is~\cite{AKM:2014}:
\begin{equation} \label{eq:Hamiltonian}
H=\hbar[-\Delta_0+g_0(b^\dagger+b)]a^\dagger a+\hbar\omega_mb^\dagger b
+\hbar\alpha_L(a^\dagger+a),
\end{equation}
where $a^\dagger$ and $a$ are the creation and annihilation operators for
the optical field, $b^\dagger$ and $b$ are the corresponding phonon 
operators for the mechanical cantilever, $\Delta_0=\omega_{d}-\omega_{cav}$ 
is the detuning between the driving laser and the optical cavity field, and 
$\omega_m$ is the resonant frequency of the mechanical mode. The quantity
$\alpha_L$ is the classical amplitude of the driving laser field, which 
is related to its power $P$ through 
$|\alpha_L|^2=2\kappa P/(\hbar\omega_{d})$, where $\kappa$ is 
the quality factor of the optical cavity. The basic physics behind the 
optomechanical coupling~\cite{L:1995} is that a change in the position 
of the cantilever, which is proportional to $(b^\dagger+b)$, can lead to 
a change in the resonant frequency of the optical field with a strength 
factor $g_0$, where $g_0\approx (\omega_{cav}/l_0)\sqrt{\hbar/(2m\omega_m)}$,
with $l_0$ being the nominal cavity length. 

\paragraph*{Calculation of classical trajectories.}
A conventional approach to investigating the dynamics of an optomechanical 
system is to use the quantum input-output theory~\cite{GZ:book} to
obtain the standard quantum Langevin equations in the Heisenberg picture.
While dissipation and fluctuations of the photon and phonon fields
have been taken into account, these are operator equations with stochastic 
fluctuations. In the classical limit ($\hbar\rightarrow 0$), i.e., bad cavity
limit, the quantum correlations between the operators 
are negligible as compared with their averages, so we have~\cite{ME:2009} 
$\langle(b^\dagger+b)a\rangle\approx\langle b^\dagger+b\rangle\langle a\rangle$.
Under this approximation, the operator equations can be replaced by those 
for the corresponding mean values, leading to the semiclassical Langevin 
equations. The deterministic dynamics of the system can be assessed by 
neglecting the small fluctuations in the photon and phonon fields. The 
resulting deterministic equations are:
\begin{equation} \label{eq:classical}
\begin{aligned}
d\langle a\rangle/dt =& i(\Delta_0\langle a\rangle
+g_0\langle a\rangle\langle b^\dagger+b\rangle
-\alpha_L)- (1/2)\kappa\langle a\rangle \\
d\langle b\rangle/dt =& - i(g_0|\langle a\rangle|^2
+\omega_m\langle b\rangle)-(\Gamma_m/2)\langle b\rangle,
\end{aligned}
\end{equation}
where $\Gamma_m$ is the dissipation rate.
A property of the classical equations is that, if 
$b$ and $a$ are replaced by $g_0b$ and $a/\alpha_L$, respectively, 
the resulting equations contain the parameter $P\propto g_0^2\alpha_L^2$,
where $g_0$ and $\alpha_L$ no longer appear as individual parameters. 
If other parameters are kept constant, the dynamics of the classical system
is solely determined by the power of the driving laser field, i.e., $P$, 
with $g_0$ and $\alpha_L$ as scaling factors. Intuitively, this can be 
understood by noting that, when a quantum system approaches its classical
limit, $\hbar$ vanishes so that the quantum strength factors
$g_0\propto\sqrt{\hbar}$ and $\alpha_L\propto1/\sqrt{\hbar}$ (both containing
$\hbar$) are degenerate into a single parameter $P$ that does not contain
$\hbar$. However, in the stochastic Langevin equations, the strengths of the
quantum fluctuations associated with the photon and phonon fields are
proportional to $g_0$ and $1/\alpha_L$, respectively. In the moderate and deep
quantum regimes away from the classical limit, as $g_0$ is increased, the
deterministic Langevin equations are less meaningful due to the more
pronounced quantum fluctuations. 

\begin{figure}
\centering
\includegraphics[width=\linewidth]{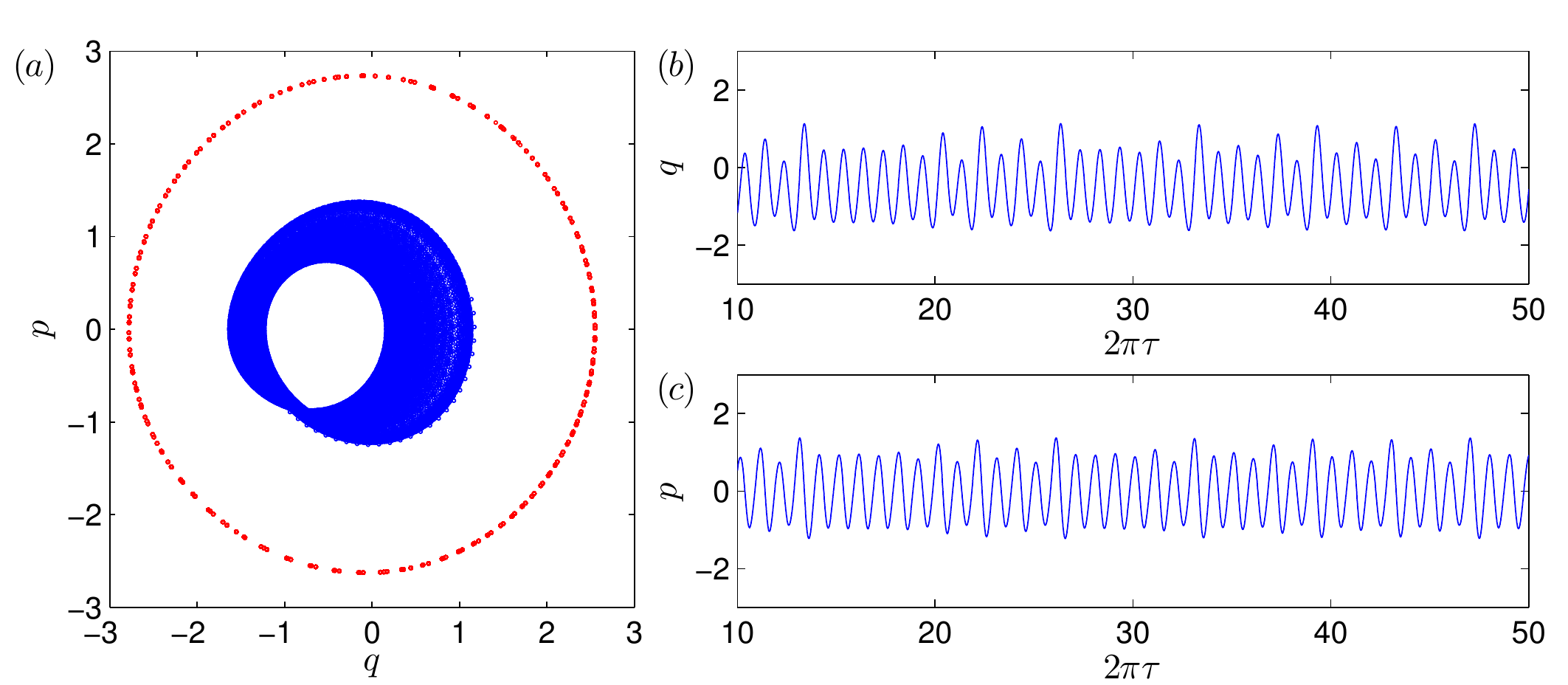}
\caption{From the deterministic classical equation, (a) a representative
chaotic trajectory and (b,c) the corresponding time series for $q$ and
$p$. The dashed circle in (a) indicates a coexisting periodic attractor.}
\label{fig:classical}
\end{figure}

The classical equations are nonlinear, so chaos can arise, as uncovered in
previous experimental~\cite{CV:2007,CCV:2007} 
and theoretical~\cite{BAF:2015,WHLG:2014,MYSXLYW:2014} works. 
To demonstrate the chaotic behavior, we use the same parameter setting as 
in the recent work of Bakemeier et al.~\cite{BAF:2015}:
$\kappa/\omega_m=1.0$, $\Gamma_m/\omega_m=10^{-3}$, 
$\Delta_0/\omega_m=-0.7$, and $\tilde{P}=8\alpha_L^2g_0^2/\omega_m^4=1.5$. 
Figure~\ref{fig:classical}(a) shows a representative 
chaotic orbit in the two-dimensional subspace of the variables 
$q=(g_0/\sqrt{2}\omega_m)\langle b^\dagger+b\rangle$ and 
$p=(-ig_0/\sqrt{2}\omega_m)\langle b^\dagger-b\rangle$, where the 
evolution time $\tau$ is made dimensionless through $\tau \equiv \omega_m t$. 
The corresponding chaotic time series is shown in 
Fig.~\ref{fig:classical}(b-c). 

\paragraph*{Calculation of quantum trajectories.}
The quantum evolution of the optomechanical system can be calculated by
using the quantum Master equation, which incorporates the effects of 
photon and phonon dissipation through the Lindblad operators. In particular,
at zero temperature the quantum Master equation is~\cite{AKM:2014,LKM:2008}
\begin{equation} \label{eq:master}
d\rho/dt = -(i/\hbar)[H,\rho]+\Gamma_m\mathcal{D}[b,\rho]
+\kappa\mathcal{D}[a,\rho]
\end{equation}
where the Lindblad operator is given by
\begin{equation} \label{eq:Lindblad} 
\mathcal{D}[L,\rho]=L\rho L^\dagger-(L^\dagger L\rho + \rho L^\dagger L)/2,
\end{equation}
and $L$ stands for either $a$ or $b$. The quantum Master equation 
describes the time evolution of an ensemble of identical quantum 
systems. The dimension of the optomechanical system is $(N_aN_b)^2$, 
where $N_a$ and $N_b$ denote the highest photon and phonon Fock states, 
respectively. An approach to reducing the dimension to $(N_aN_b)$ is 
to ``unravel'' the deterministic quantum Master equation through the 
stochastic wavefunction equation for quantum 
trajectories~\cite{GP:1992,GP:1993a,GP:1993b,SB:1997}. The deterministic 
property is retained through the ensemble average of many realizations 
of the system starting from the same initial condition. Among the many
unraveling schemes for generating quantum trajectories, the QSD
approach is convenient and efficient with results that can be related to
the record of homodyne detection, an important measurement tool in 
optomechanics~\cite{WHHRHA:2015}. The QSD equation is
given by~\cite{DGS:1998,SDG:1999}
\begin{equation}
    \label{eq:OMS_QSD}
\begin{split}
|d\psi\rangle =&-(i/\hbar)H|\psi\rangle dt + 
\sqrt{\kappa}(a-\langle a\rangle)
|\psi\rangle\circ(d\xi_1+\sqrt{\kappa}\langle a^\dagger\rangle dt)
-(1/2)\kappa(a^\dagger a-\langle a^\dagger a\rangle)|\psi\rangle
dt \\
&+\sqrt{\Gamma_M}(b-\langle
b\rangle)|\psi\rangle\circ(d\xi_2+\sqrt{\Gamma_M}\langle b^\dagger\rangle dt)
 - (1/2)\Gamma_M(b^\dagger b-\langle b^\dagger b\rangle)|\psi\rangle dt,
\end{split}
\end{equation}
where $\langle O\rangle=\langle\psi|O|\psi\rangle$ is the expectation value 
of operator $O$ for the specific wave function $|\psi\rangle$. 
The QSD equation (\ref{eq:OMS_QSD}) is in fact a Stratonovich type of 
stochastic equations. (The Ito form of QSD has also been established 
and widely used~\cite{GP:1992,GP:1993a,GP:1993b,B:2000,BPS:1996}.) 
In the QSD equation, the terms $d\xi_j$ $(j=1,2)$ are complex 
Gaussian white noise for the photon and phonon fluctuations, which satisfy 
\begin{equation*}
Md\xi_j=Md\xi_id\xi_j=0 \ \ \mbox{and} \ \ Md\xi_i^*d\xi_j=\delta_{ij}dt,
\end{equation*}
where $M$ stands for the ensemble average. The density operator can be 
reconstructed through the mean over the projectors of the ensemble 
quantum states 
\begin{equation} \label{eq.rho}
\hat{\rho}=M|\psi\rangle\langle\psi|.
\end{equation}

\begin{figure}
\centering
\includegraphics[width=\linewidth]{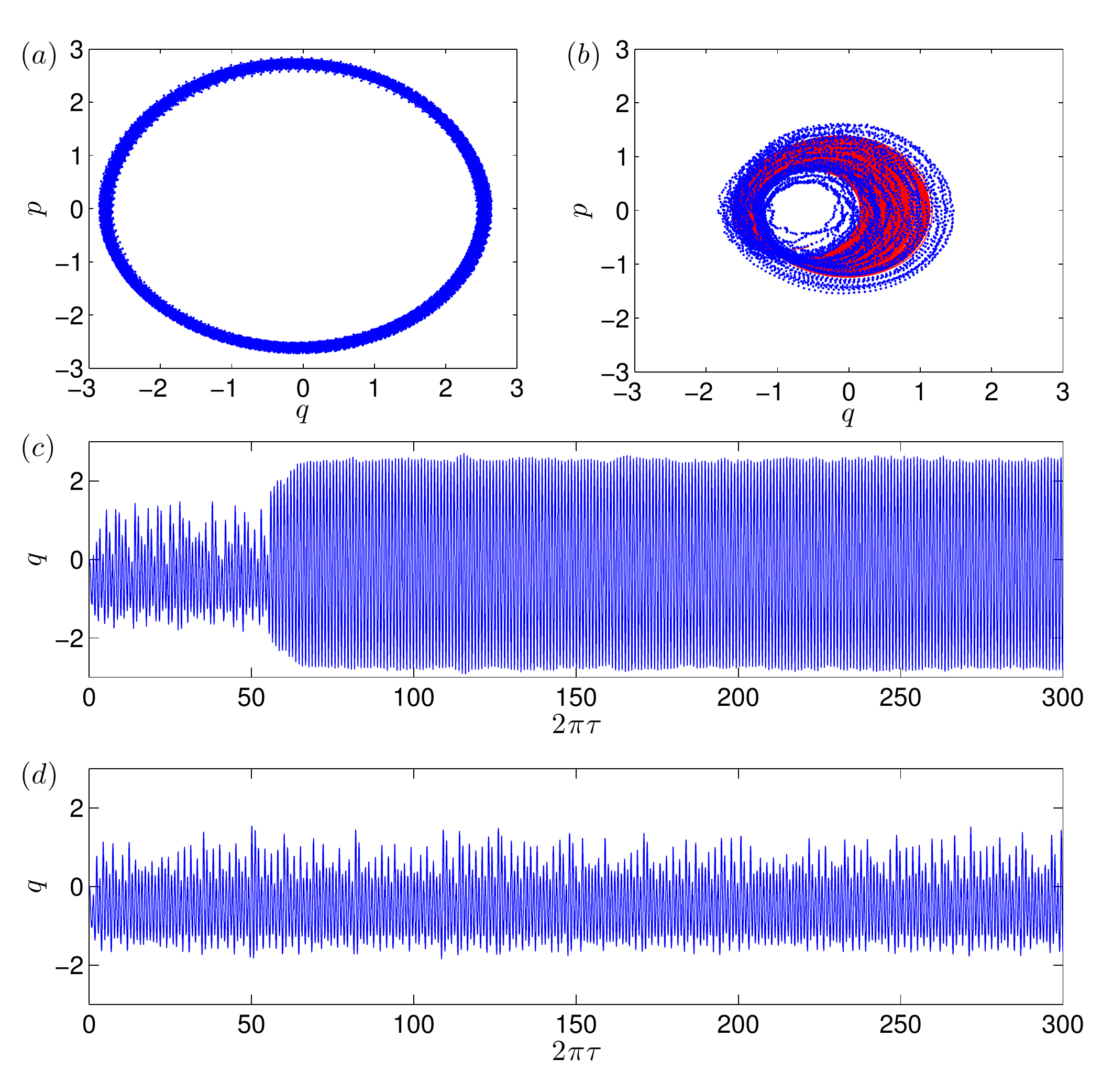}
\caption{For $g_0/\omega_m=0.1$, (a) an asymptotic quantum
trajectory calculated from the QSD method, (b) the quantum trajectory in the
transient phase, overlapped with the corresponding classical trajectory,
(c) the corresponding time series. The asymptotic quantum trajectory is
regular, in spite of the quantum fluctuations. However, the transient
quantum trajectory is chaotic and coincides well with the classical
trajectory (gray). (d) An example of a very long chaotic transient in
the quantum regime for $g_0/\omega_m=0.05$.}
\label{fig:quantum_1}
\end{figure}

In an optomechanical system, the quantum effects can be characterized by 
the parameters $g_0\propto\sqrt{\hbar}$ and $\alpha_L\propto 1/\sqrt{\hbar}$. 
Figure~\ref{fig:quantum_1}(a) shows, for $g_0/\omega_m = 0.1$, a typical 
quantum trajectory calculated from the QSD equation in the $(q,p)$ plane. 
This is a periodic, limit-cycle trajectory, despite being noisy due to 
the quantum fluctuations. The corresponding time series $q(2\pi\tau)$ 
is shown in Fig.~\ref{fig:quantum_1}(c). Since the value of
the laser power $P$ is fixed, the corresponding classical behavior is 
that shown in Fig.~\ref{fig:classical}, which is chaotic. The remarkable
phenomenon is that, the quantum trajectory in Fig.~\ref{fig:quantum_1}(a)
is characteristically different from the classical trajectory in 
Fig.~\ref{fig:classical}(a): the former is regular while the latter is 
chaotic! This is the recently discovered phenomenon of quantum-classical 
correspondence breakdown in optomechanical systems~\cite{BAF:2015}. 

\paragraph*{Transient chaos in the quantum regime.}
We find that the breakdown can be naturally viewed as a manifestation 
of transient chaos. We note from Fig.~\ref{fig:quantum_1}(c) that, before 
the periodic quantum state is reached, there is a relatively short time 
interval during which the quantum evolution is characteristically different, 
which is a transient phase. The quantum trajectory of the system in the 
transient phase is shown in Fig.~\ref{fig:quantum_1}(b), which appears chaotic. 
The striking finding is that, the transient quantum trajectory is remarkably 
consistent (in fact coincides) with the corresponding classical trajectory 
(the red background trajectory in Fig.~\ref{fig:quantum_1}(b), which 
overlaps with the quantum trajectory almost completely). As we tune the 
parameter $g_0/\omega_m$ towards the classical regime, the duration of the 
transient phase increases. The extreme situation is that the transient
time becomes so long that the system stays in a chaotic state for any 
practical time. An example is shown in Fig.~\ref{fig:quantum_1}(d) 
for $g_0/\omega_m = 0.05$. 


\begin{figure}
\centering
\includegraphics[width=\linewidth]{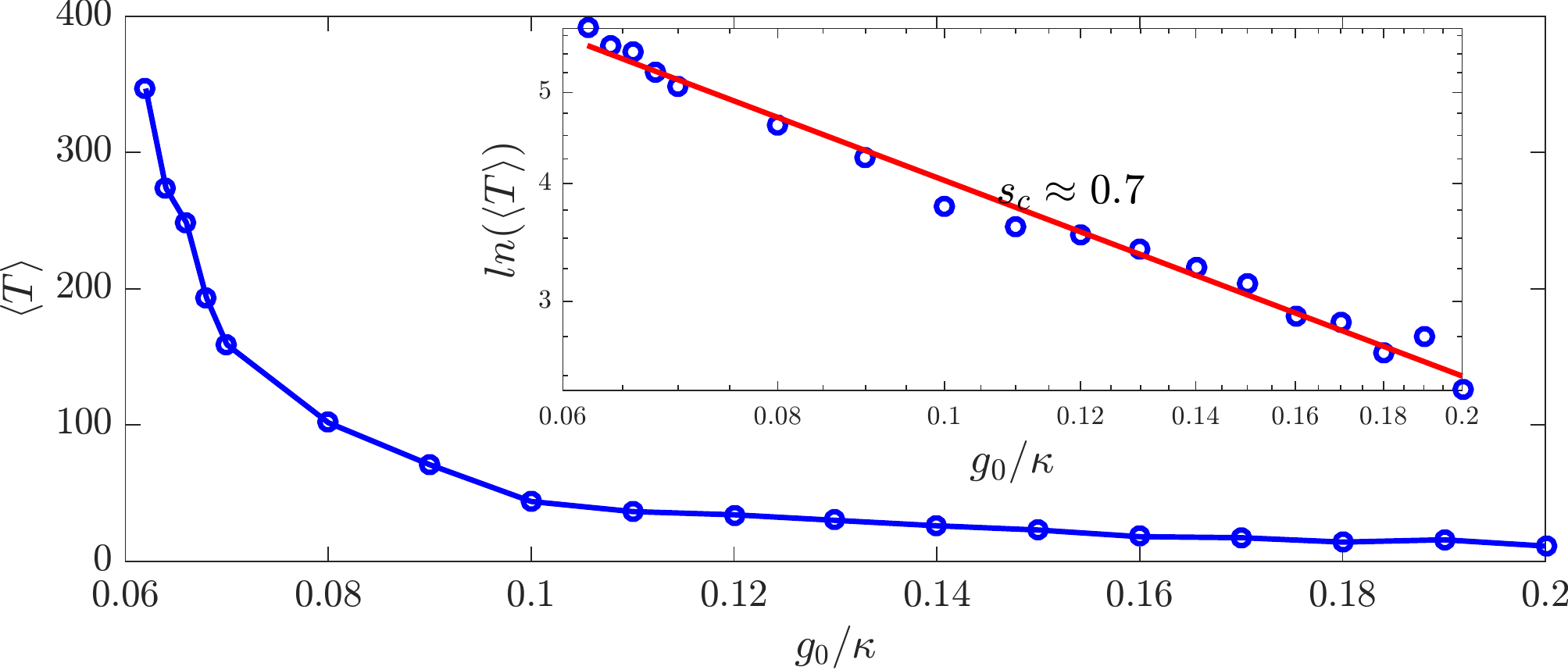}
\caption{QSD Results: Dependence of average chaotic transient lifetime,
$\langle T\rangle$, on $g_0$ on a linear-linear plot and on a double
logarithmic versus logarithmic scale (inset). All points are result of
averaging 100 QSD realizations.}
\label{fig:scaling}
\end{figure}

How does the average chaotic transient lifetime $\langle T\rangle$ depend on the
quantum strength parameter $g_0$? 
Here, the quantity $\langle T\rangle$ is the average time required for the
system to transition from a chaotic attractor in the classical limit
to a coexisting periodic attractor in the quantum regime, as induced
by quantum fluctuations. For example, for a specific trajectory in
Fig.~3(c), the transition occurs at $2\pi\tau\approx55$, so the
transition time is $T=55/2\pi$. (Note that the evolution time
$\tau$ is made dimensionless through $\tau\equiv\omega_mt$). From
Fig.~3(c), we also see a dramatic change in the amplitude before and
after the transition, and this can be exploited for efficiently computing the  
average transition time from a large number of quantum trajectories.
In general, the time from the beginning to the end of the transition can be
neglected as compared with the typically long transient time,
especially when the effective Planck constant is reduced.
As shown in Fig.~\ref{fig:scaling}, as $g_0$ is decreased so that 
the quantum effect becomes progressively weaker, $\langle T\rangle$ 
increases dramatically. A qualitative
explanation for Fig.~\ref{fig:scaling} is the following. The classical
trajectories are calculated from the deterministic, semiclassical Langevin
equation in the Heisenberg picture with dissipation, where quantum fluctuations
are neglected. The quantum trajectories are obtained from the QSD method, an
unraveling of the general quantum Master equation in the Schr\"{o}dinger picture
using the Lindblad operators.  The quantum fluctuations in QSD not only play the
role of noise in the classical deterministic system, but more importantly, they
can induce characteristic changes in the system dynamics. Say we fix the laser
power so that the classical dynamics remains chaotic. What will happen when the
quantum effects (fluctuations) become increasingly pronounced? Mathematically,
as $g_0$ is increased, it is necessary to decrease $\alpha_L$ to keep the
driving laser power constant. This effectively enhances the ratios
$\sqrt{\kappa}/\alpha_L$ and $\sqrt{\Gamma_m}/\alpha_L$ in the QSD equation,
which are the relative noise-to-driving ratios. As noise becomes more
pronounced, the probability that the system can stay in the deterministic
chaotic set is decreased, reducing the chaotic transient lifetime.

To further test the proposition that noise or quantum fluctuations can drive
the quantum system away from the classical chaotic invariant set, we calculate  
the quantum trajectories but with the noise term excluded. We find that, without
random fluctuations, the quantum trajectories follow the classical chaotic 
set {\em all the time}. This result confirms that it is the quantum 
fluctuations which eventually drive the quantum trajectories out of the
classical chaotic set, generating transient chaos. The weaker the quantum
fluctuations, the longer the average transient lifetime will be. The 
quantum-classical transition is thus induced by quantum fluctuations,
which resembles the phenomenon of noise-induced transition in classical
systems that can be treated using the classical Kramer rate 
theory~\cite{HTB:1990}. The transient chaos associated with 
quantum-classical transition is also relevant to the quantum activation 
process~\cite{Dykman:2007}, a transition process induced by noise between
coexisting asymptotic states in a quantum system. 
We remark that, in a related work~\cite{C:2012}, it was reported that
quantum isoperiodic stable structures can be retained by the information 
from the classical isoperiodic stable structures in presence of noise.

\paragraph*{Scaling of transient lifetime and physical understanding.}
The Kramer theory or the quantum activation theory stipulates that the 
escape rate $\kappa$ generally follows the scaling as 
\begin{displaymath}
\kappa = \nu exp(-E_b/E_{noise}), 
\end{displaymath}
where $E_b$ denotes the threshold energy for activation, $\nu$ is a 
prefactor, and $E_{noise}$ is the strength of the fluctuation, e.g.,
on the order of $k_B\tilde{T}$ due to the thermal environment or
$\hbar\omega$ in the deep quantum regime, where $\tilde{T}$ represents 
temperature. At low temperatures, the quantum fluctuations are dominated 
by the zero-point energy.

Figure~\ref{fig:scaling} shows the relation between the average chaotic 
transient lifetime $\langle T\rangle$ and the magnitude $g_0$ of the 
quantum fluctuations on a double logarithmic scale. The relation can be 
well fitted by a straight line, as shown in the inset of 
Fig.~\ref{fig:scaling}, which indicates the scaling law:
\begin{equation} \label{eq:scaling}
\ln{\langle T\rangle} \sim (g_0/\kappa)^{-s},
\end{equation}
where $-s$ ($s>0$) is the slope of the linear fit. The scaling law is 
characteristic of superpersistent chaotic transients in nonlinear 
dynamical systems~\cite{GOY:1983,GOY:1985,CK:1988,LW:1995,DL:2003}.
The physical meaning is that, as the quantum fluctuations are reduced
so that the classical description becomes more accurate, the chaotic 
behavior becomes significantly more persistent in that its lifetime 
increases faster than the Ehrenfest time. 

\begin{figure}
\centering
\includegraphics[width=\linewidth]{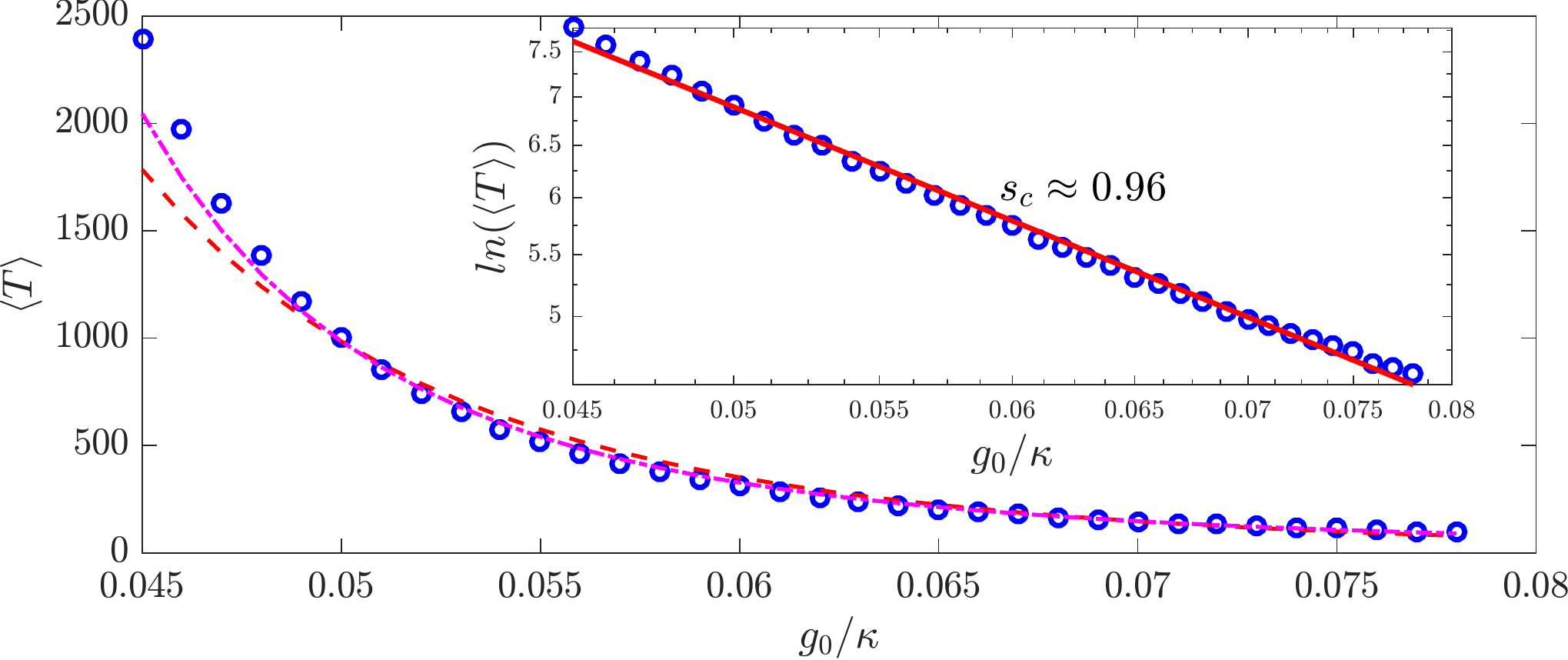}
\caption{Results from classical Langevin equation: (a) Dependence of
the average chaotic transient lifetime, $\langle T\rangle$, on $g_0$ on
a linear-linear plot. Inset: the same plot but on a double logarithmic
versus logarithmic scale. The magenta dash-dot curve is a fit of the
superpersistent chaotic transients behavior while the red dash curve
is a fit of the Ehrenfest scaling. In the inset the red straight curve
shows the slope of the supperpersistent chaotic transients behavior is
about $-s_c\approx -0.96$. All points are result of averaging 10000
Langevin equation realizations.}
\label{fig:cla_scaling}
\end{figure}

To better understand the scaling behavior of the average transient
lifetime, we exploit the quantum Langevin equations:
\begin{equation}
\begin{split}
dq/d\tau=&p, \\
dp/d\tau=&(\sqrt{2}/8)\tilde{P}|\alpha|^2-q-\Gamma_Mp+g_0\xi, \\
d\alpha/d\tau=&i(\Delta_0\alpha+\sqrt{2}\alpha q-1)-\kappa
\alpha/2+(\sqrt{\kappa/2}/\alpha_L)\alpha_{in}. 
\end{split}
\end{equation}
In general, a Langevin equation can be analyzed using the corresponding
Fokker-Planck equation, where the stochastic component of the former 
contributes to the diffusion term in the evolution of the probability
distribution of the latter.
For the Fokker-Planck equation, a general solution cannot be written
down explicitly except for one-dimensional systems. In this case, the steady
state distribution has the form $W_s(x)=\mathcal{N}\exp{[-U(x)/D]}$, where
$U(x)$ is the effective potential, $D$ is the noise amplitude proportional 
to $g_0^2$, and $\mathcal{N}$ is a normalization constant. The mean
first passage time over a barrier, i.e., the diffusion time from a local 
minimum $U(a)$ over a saddle point $U(b)$, obeys the following scaling
law~\cite{Gardiner:book} with $D$: $T_{MFP}\propto e^{[U(b)-U(a)]/D}$. 
However, to predict the exact form of the scaling law from the general 
multivariable Fokker-Plack equation is difficult. An alternative is to
calculate the average chaotic transition lifetime (or the mean first
passage time) from the Langevin equations. The results are shown in
Fig.~\ref{fig:cla_scaling}. Due to the relative simplicity of the Langevin
equation as compared with the QSD equation, it is possible to probe more
deeply into the classical regime with much longer transition lifetime. 
We find that, in the $g_0$ regime where both types of results are available,
the agreement is excellent. In particular, solutions of the Langevin
equation gives
\begin{displaymath}
\ln{\ln{\langle T\rangle}}/\ln{g_0} \approx -1.
\end{displaymath}
In Fig.~\ref{fig:cla_scaling}(a), we show the fitting curve of the
Ehrenfest scaling (red dash) as well as the superpersistent chaotic transition
behavior (magenta dash-dot). For the Ehrenfest scaling, we use the 
least-squares method to fit $\langle T\rangle=C_0\cdot g^{-\delta}$ on a 
double logarithmic scale. For the superpersistent scaling, it is not 
straightforward to fit the relation $\langle T\rangle=C_1\cdot e^{C_2/g^s}$. 
We thus set $C_1=1$ and fit the simulation results in terms of 
$\log{[\log{(\langle T\rangle)}]}$ versus $\log{g_0}$. We
see that the magenta curve fits better than the red curve, especially in 
the middle region. For small values of $g_0/\omega_0$, the Ehrenfest 
scaling exhibits larger deviations from the simulation results as compared 
with the superpersistent transient scaling.

For the QSD results (Fig.~\ref{fig:scaling}), we estimate the slope of the 
fitting line of $\ln{\ln{(\langle T\rangle)}}$ with $\ln{g_0}$ and obtain the 
absolute value of about $0.7$, which is smaller than the result from 
the Langevin equation. There can be multiple reasons for the difference.
For example, for a large value of $g_0/\omega_0$, the trajectories tend to 
approach the periodic attractor from the beginning. However, the transition 
process takes time, so the state at an arbitrary instant of time during 
the transition is actually recorded. When the transition time is comparable
with the transient time, error can occur. Considering that our system is
higher than one dimensional and the simulations were done with the full quantum
state diffusion equation, the difference in the slope may not be unreasonable.
In particular, in high dimensions the slope should have a smaller absolute value
because of the existence of more ``paths'' to cross the saddle point (there is
only one route in one dimension), facilitating the transition. 

\begin{figure}
\centering
\includegraphics[width=\linewidth]{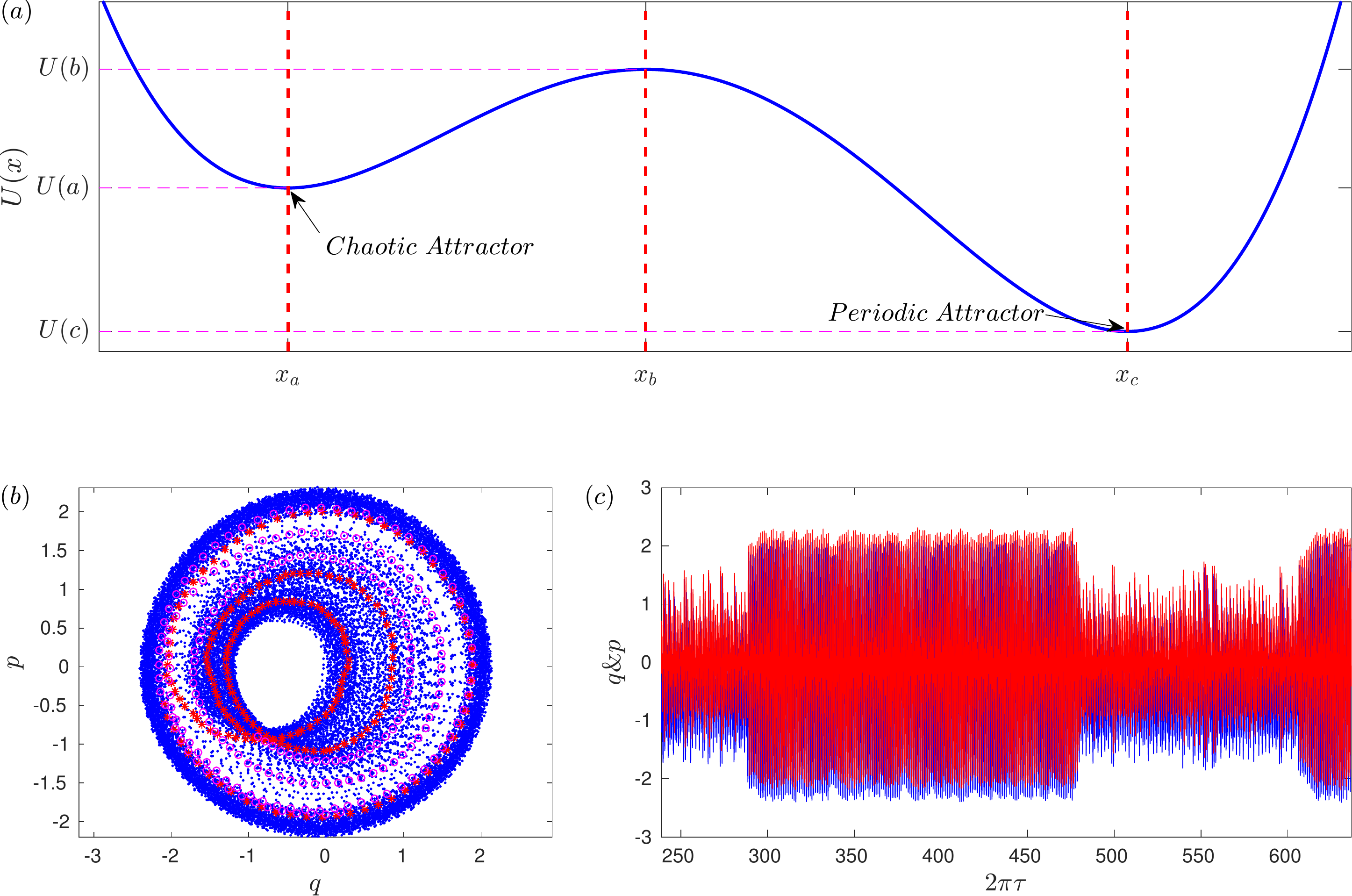}
\caption{(a) A mechanical picture illustrating the noise-induced
transition between chaotic and periodic attractors, where the periodic
attractor is more stable than the chaotic attractor. For $g_0/\omega_m=0.056$,
representation in the $q-p$ space (b), where the red stars represent the
transition process from the inner chaotic attractor to the outer periodic
attractor while the magenta circles represent the transition in the opposite
direction. (c) The corresponding time series, where the blue and red colors
are for $q$ and $p$, respectively.}
\label{fig:backwardtran}
\end{figure}

A natural question is whether the reverse process, i.e., transition from the
periodic orbit to the chaotic orbit, can happen. In nonlinear dynamics, periodic
attractors are usually more stable than  chaotic attractors. Heuristically, a
system in which a periodic and a chaotic attractors coexist can be viewed as
particle motion in a mechanical system with two asymmetric potential wells
subject to unbounded (e.g., Gaussian) noise, where the periodic attractor
corresponds to the deep well and the chaotic attractor is associated with the
shallow well, as schematically shown in Fig.~\ref{fig:backwardtran}(a). The
probability for the particle to  ``hop'' into the shallow well from the deep
well is considerably smaller than that in the opposite direction. In
optomechanical systems, this kind of backward transition can occur but it is
rare. One such case is shown in Figs.~\ref{fig:backwardtran}(b) and
\ref{fig:backwardtran}(c), where the transition occurs at $g_0/\omega_0=0.056$.
For smaller values of $g_0$, it is highly unlikely that the trajectory can
switch into the periodic attractor. Even if this occurs, the probability for the
trajectory to escape the periodic attractor will be exponentially small due to
the higher potential barrier. For large values of $g_0$, transition in both
directions can occur, as shown in Fig.~\ref{fig:backwardtran}(c).

Our reasoning based on separating the deterministic and stochastic
components of the Langevin equation does not depend on the specific 
details of the system, suggesting that the fast growing behavior in 
the average transient lifetime and the associated scaling law are 
{\em generic}.  

\section*{Discussion}

To summarize, we investigate the fundamental problem of quantum-classical
correspondence in optomechanical systems from the perspective of dynamical
evolution. When the classical system exhibits chaos, the evolution of the 
quantum system contains two phases: chaotic motion in the (relatively) short 
time scale and regular motion in the long time scale. The transient chaotic 
behavior of the quantum system corresponds precisely to that in the classical 
limit - in this sense there is a well-defined quantum-classical 
correspondence. The long term behavior of the quantum system, however, is 
characteristically different from the classical behavior - in this sense 
there is a breakdown~\cite{BAF:2015} of the quantum-classical correspondence. 
As the classical regime is approached, the chaotic transient lifetime 
increases dramatically (faster than the Ehrenfest time for isolated 
systems - see below). Our finding of transient chaos in optomechanical 
systems, besides being a remarkable phenomenon by itself, provides a natural 
resolution for the paradoxical breakdown of quantum-classical correspondence.

In general, the problem of quantum-classical correspondence can be addressed 
through the approach of quantum-classical transition (QCT). It is known that, 
unlike special relativity where Einstein's theory can be smoothly transformed 
to Newtonian mechanics in the limit $v/c \rightarrow 0$, the approach 
of a quantum system to the classical limit $\hbar\rightarrow0$ is singular. 
In the classical world, chaos exists in both dissipative and Hamiltonian
systems, and chaotic dynamics are often studied in the phase space. 
However, to our knowledge, attempts to find chaos in the Schr\"{o}dinger 
equation or in the quantum Liouville equation have not been convincingly
successful. One reason is that isolated quantum systems are fundamentally 
linear. Another reason is that, the uncertainty principle forbids 
arbitrarily fine scale structures in the phase space. Indeed, in bounded 
and isolated (or closed) quantum systems the most complicated dynamics are 
quasiperiodic. Even though the transient behavior of a quantum system 
can be similar to that in the corresponding classical system, any classical
features will be lost after a time scale called the Ehrenfest time: 
$t_E\propto\hbar^{-\delta}$, where $\delta$ is determined by the details of 
the system. Strictly, the Ehrenfest time holds for the idealized situation
where the underlying system is fully closed. With the development of the 
quantum theory and advances in experimental techniques, the quantum 
dynamics of other types of situations have been considered, such as
{\it unconditioned open} and {\it conditioned open} 
systems~\cite{H:book,BHJ:2000}. In the former case, the system is coupled
to the environment but no information about the system is extracted,
while for the latter information about the state of the system is
extracted from it. For an unconditioned open system, the dynamical evolution 
is governed by the quantum Master equation, which is still linear. However,
for a conditioned open system, its dynamical evolution follows a stochastic 
quantum equation that contains a nonlinear term representing the conditioning 
due to the measurement.

In the study of QCT, there are two general approaches to addressing the
quantum-classical correspondence. The first 
is to focus on the agreement between the distribution functions, i.e., 
the quantum Wigner and the classical distribution functions - 
the weak form of QCT~\cite{KLRS:2008,KRSL:2007}. The second approach,
the strong form of the QCT~\cite{BHJ:2000,HJS:2006}, is to examine the 
localization of the quantum trajectory on the classical orbits, in which 
chaos can emerge naturally. To assess the degree of localization,
continuous measurements of the system are required, introducing a 
nonlinear term in the quantum equation, so this approach is applicable
only to conditioned open systems.

\begin{table}[t]
\parbox{15.0cm}{
\caption{\label{tab:1}
An overview of distinct QCT regimes.}} \\
\centering
\begin{tabular}{|c|c|c|c|} \hline
       & Conventional QCT & Weak QCT & Strong QCT \\ \hline
      System & Isolated & Unconditioned Open & Conditioned Open \\ \hline
      Equation & Schr\"{o}dinger & Master & Quantum Trajectory \\ \hline
      Dynamics & Linear & Linear & Stochastic Nonlinear \\ \hline
      Characteristic Time & Ehrenfest time $\sim\hbar^{-\delta}$ &
      Unknown at present  & {\em Eq.~\eqref{eq:scaling}
      discovered in this paper}  \\ \hline
\end{tabular}
\end{table}
   
Table~\ref{tab:1} presents an overview of the status of the knowledge
about QCT, with knowns and unknowns specified. An outstanding issue 
concerns the scaling of the transition time in the strong QCT regime.
In particular, the question is whether the QCT time follows the same scaling 
law as the Ehrenfest time. We address this issue in this paper by exploiting
optomechanical systems subject to continuous heterodyne detection, which
fundamentally exhibits a strong form of QCT. Qualitatively, our main 
finding is that transient chaos effectively serves as a bridge for the 
QCT. Quantitatively, we uncover a scaling law for the transition time
which is different from that for the Ehrenfest time associated with 
the conventional QCT for isolated systems. With the advances in 
experimental techniques, there is now ability to observe quantum 
trajectories~\cite{BAF:2015,CSBSMRH:2016}. We expect the main
results of this paper to be experimentally testable.  

We make a few further remarks pertinent to our results.

\paragraph*{Remark 1: Transient chaos in quantum systems - what
does it mean?} A quantum system is fundamentally linear. How can then a
quantum trajectory be chaotic, even transiently? This paradox can be 
resolved, as follows. The Schr\"{o}dinger (quantum Master) equation
describes the time evolution of an individual system in an ensemble
of identical systems, from which the mean value of any physical quantity,
$\langle\psi|\hat{O}|\phi\rangle$
[$\mbox{Tr}(\hat{O}\rho)$] for an operator $\hat{O}$, can be obtained.
This is an ideal evolution process during which no further
disturbance or measurement should be made; for otherwise the wavefunction
will collapse into an eigenstate determined by the whole system, including
the measurement apparatus. Based on the quantum trajectory theory,
the time evolution of the mean value can also be produced
from the QSD calculations through the ensemble average. When there is chaos
in the classical limit, the ensemble averaging process can make the time
evolution of the mean value periodic. That is, even when QSD calculation gives
that a single quantum trajectory is chaotic in the transient phase, the
ensemble average of many such trajectories can still be regular. 

\paragraph*{Remark 2: Effect of measurement.}
A single quantum trajectory, however, is not physically 
meaningless. For a dissipative quantum system, the Master equation can 
yield the best prediction about the dynamical evolution of an ensemble 
of the system in absence of any measurement. The single trajectory 
calculated from quantum methods, such as QSD and quantum jump 
theory, has the physical meaning of conditioned realization of an individual
system under a particular observation record, through homodyne/heterodyne
detection and photodetector~\cite{PK:1998,W:1996}. This makes the quantum
trajectories {\it subjectively} real~\cite{W:1996}. In the continuously
conditioned measurement theory, any measurement introduces a factor called 
the detector efficiency, $0 \le \eta \le 1$, into the QSD
simulation~\cite{WM:1993a,WM:1993b,DJ:1999,BG:2003}, which models
the situation where the beam-splitter transmittivity is less than
unity~\cite{WM:1993a}. Mathematically, this factor can be taken into 
account by decomposing the photon fluctuation term into two uncorrelated 
terms of strength $\sqrt{\eta}$ and $\sqrt{1-\eta}$,
respectively~\cite{WM:1993a,BG:2003}. In the limit $\eta \rightarrow 1$, this
form of stochastic equation is reduced to the equation simulated in our work,
which corresponds to perfect detection. 

\paragraph*{Remark 3: Effect of temperature.}
Our treatment of the breakdown of quantum-classical correspondence in 
optomechanical systems as a problem of strong QCT assumes the low 
temperature limit. As we argue and demonstrate, the transition from chaos
to a regular state is mediated by quantum fluctuations or noise. Naturally
we expect that thermal noise would play a similar role. In particular, if
we focus on the classical system subject to thermal noise, a transition 
from a chaotic attractor to a periodic one can occur, accompanied by
transient chaos. 

In an optomechanical system, the fundamental physical constant $\hbar$
cannot be changed in experiments. The degree of quantum fluctuations can
be controlled by adjusting or engineering other parameters, e.g., the mass
of the cantilever, while keeping the system at low temperature. Weaker 
quantum fluctuations corresponding to smaller values of $g_0$ can be 
realized using a heavier cantilever. In this case, the quantum system 
would behave chaotically for a relatively long time, due to the exponentially
long transient lifetime. We note that, in the high temperature regime,  
the strength of the quantum fluctuations scales as $\sim g_0^2(2\bar{n}+1)$, 
where $\bar{n} \sim  k_bT/\hbar\omega_m$. This indicates a 
counter-balancing effect between temperature $T$ and mass $m$ as $\sim T/m$. 
Consequently, at high temperatures a system of relatively large mass can 
still behave chaotically for a long time. 

\paragraph*{Remark 4: Emergence of chaos in the quantum regime.}
In the works of Habib et al.~\cite{GADBH:2004,HJS:2006}, the occurrence 
of chaos in the quantum regime was reported. The underlying mechanism
lies in continuous measurement, which is key to resolving the 
quantum-classical correspondence, as demonstrated in our work.
 
A pertinent question is, since the transition time from classical chaotic 
to quantum regular motions can be quite short, why would there be chaos in 
the deep quantum regime as studied by Habib et al.? In these works, a toy 
model was studied and the main idea was to vary the effective Planck 
constant $\hbar$ (e.g., from $10^{-2}$ to $16$) and the measurement 
strength $k$ to calculate the quantum Lyapunov exponent $\lambda$
and compare its values with those of the classical exponent $\lambda_{Cl}$.
It was found~\cite{GADBH:2004,HJS:2006} that, for $\hbar=10^{-2}$, the 
classical results can be recovered through continuous changes in the 
measurement strength. Furthermore, non-negative values of $\lambda$ were 
obtained before the classical limit, indicating that chaos may exist in the 
regime far away from the classical limit. 

The optomechanical systems we studied are experimentally realizable, for
which there are realistic criteria to determine if the system is in a
classical or in a quantum regime. In particular, to characterize an
optomechanical system, a number of key dimensionless parameters can be used
- see, e.g., Eq.~(119) in Ref.~\cite{AKM:2014}.
The most relevant parameter is $g_0/\kappa$ - the
``quantumness'' parameter, where $g_0/\kappa>1$ represents the strong couping
regime in which the optical device can detect the change of even one
phonon. In our work, we used $g_0/\kappa$ in the range $10^{-2}\sim10^{-1}$.
To determine whether this regime is quantum, we refer to Fig.~9 
in Ref.~\cite{AKM:2014}, which summarizes the
experimental parameters for various quantum realizations.
For example, in Ref.~\cite{CASHKGAP:2011}, the parameter
values $g_0/2\pi=910$ kHz and $\kappa/2\pi=500$ MHz were used, which lead to
$g_0/\kappa=1.82\times10^{-3}$. In Ref.~\cite{VDWSK:2012}, 
the values $g_0/2\pi=3.4$ kHz and $\kappa/2\pi\sim 1$ MHz were used,
giving rise to $g_0/\kappa\sim10^{-3}$. In addition, in experimental
studies of conditioned measurement of optomechanical systems~\cite{WHHRHA:2015},
the value of $g_0/\kappa$ used was about $10^{-5}$ but contributions from 
quantum noise were also taken into account. (In these experimental works, the aim 
was not finding chaos in the quantum regime.) Referring to the values of 
$g_0/\kappa$ realized in these experiments, we see that the systems studied 
in our work are in the quantum regime even for e.g., $g_0/\kappa=0.045$.
Thus, there is no conflict between our result and that of 
Habib et al.~\cite{GADBH:2004,HJS:2006}.
Considering the fact that many quantum trajectories in our system localize
on the classical chaotic attractor for thousands of periods and this time
can be made significantly longer through small changes in the the parameter
$g_0/\kappa$, from the point of view of experiments, there is chaos in
the quantum regime in our system. However, we emphasize that the main point 
of our work is not that we find chaos in the quantum regime. Our goal is to 
address the issue of quantum to classical transition quantitatively through 
a scaling analysis of the transition time. It is for this purpose that we use the
notions of ``classical limit'' versus ``quantum regime.'' 
 
\paragraph*{Remark 5: appearance of chaos in the quantum regime in absence of
classical chaos.}
There were recent reports of emergence of chaos in the quantum regime in 
absence of classical chaos~\cite{PDMLHAKP:2016,EHC:2016}. In these works,
the quantum versus classical ``weights'' of the system is controlled by
the effective Planck constant, where the classical limit is reached
when the constant approaches zero. In our system, the parameter $g_0$ plays 
the same role. In classical nonlinear dynamical systems, chaos is common and 
noise can induce transition among different attractors - these phenomena are 
usually system and parameter dependent. A main point of our work is that quantum
fluctuations can effectively serve as noise and induce transitions, with
transient time depending on the fluctuation strength. We note that B.~Pokharel 
et al.~\cite{PDMLHAKP:2016} used quantum tunneling to explain their results. 
In our work we focused on the case where the classical limit is chaotic,
and we observe transitions in both directions: from chaotic to regular
motions and vice versa, and we argue that the transition probabilities
in the opposite directions can be drastically different. In general, even
for a set of parameters for which the asymptotic classical dynamics is
regular, there can be transient chaos in relatively short time scales
due to nonattracting chaotic invariant sets. When there is noise, there
can be transitions between the regular attractor and the nonattracting
chaotic set, leading to a combined chaotic attractor. In a general sense,
quantum tunneling can induce transitions among different states and, as
a result, chaos in the quantum regime can occur in open systems subject to
continuous measurement. In this sense, our work does not contradict that
of B.~Pokharel et al.~\cite{PDMLHAKP:2016}.

\paragraph*{Remark 6: ``suppression'' of classical chaos.}
Equation~(2) holds in classical limit for which quantum fluctuations 
do not exist. In the quantum regime, the
fluctuations are naturally incorporated into the quantum trajectory
calculations. To account for the quantum fluctuations in the semiclassical
theory, we use the quantum Langevin equation~\cite{LKM:2008}. 
Note that Eq.~(8) is a set of rescaled equations so that the
fluctuation or ``noise'' strength is nothing but $g_0$. However,
the noisy version of Eq.~(2) can be studied so as to reveal the equivalence
between the effects of classical noise and quantum fluctuations.
We focus on the noise-to-driving ratio, a quantity that
increases with $g_0$. In the quantum Duffing oscillator model~\cite{EHC:2016}, 
when the parameter $\beta$ is increased, the amplitude
of the driving $(g/\beta)\cos{(\omega t)}$ is reduced. For very large
value of $\beta$, the quantum Lyapunov exponent becomes negative while
the exponent in the classical limit remains positive, indicating a
transition from chaos to a periodic behavior. We observed similar results
in our optomechanical systems, i.e., the quantum fluctuations can suppress
classical chaos.

Mathematically, suppression of classical chaos can be treated as a phenomenon
of quantum fluctuation induced transition. 
In our optomechanical system the periodic attractor is apparently more
stable than the chaotic attractor, which also appears to be the case in
the quantum Duffing system studied by J.~K.~Eastman et al.~\cite{EHC:2016} in
the parameter regime where there is chaos in the classical limit. For
small values of $g_0$ where the quantum fluctuations are weak, the classical
chaotic behavior can last for a long time, as quantified by the scaling
law uncovered in our paper. For relatively large values of $g_0$, the
transition time from chaos to a periodic behavior becomes significantly
shorter. In our paper we also discuss the reverse transition and point
out that the probability is negligibly small, as shown in Figs.~6(b,c).
While the reverse transition can occur with a larger probability for very
strong noise, in optomechanical systems such strong noise cannot be
realized with quantum fluctuations only~\cite{VGFBTGVZA:2007,ME:2009,QCHM:2012}.

\section*{Methods} 

Historically, the method of quantum trajectory represented an efficient 
way to solve the master equation, and certain types of quantum trajectories 
can correspond to the result of conditioned 
measurement~\cite{CSBSMRH:2016,GKGBDHBRH:2007,YZSWDCWH:2008,VLMZFBA:2010,
NBSRFHWJ:2010,VSS:2011,MWMS:2013,HSMSGBSAFG:2013,VPSAWBGSHC:2014,FMDS:2014,
WCDJMS:2014,LRTETJWSD:2014}. Mathematically, an ensemble of quantum systems
whose state vectors are governed by a stochastic differential equation can 
have a density operator that satisfies a unique deterministic master equation.
In contrast, a specific master equation can correspond to many different 
stochastic equations or different unravellings such as the QSD equation, 
the quantum jump equation, or the orthogonal jump equation~\cite{P:book}.
While all the unravellings can be used to simulate the master equation, 
they have a different physical meaning. The most commonly calculated quantum 
trajectories are those from the QSD equation and the quantum jump equation, 
corresponding to homodyne and photon counting detection, respectively.

For a general Lindblad form of the master equation:
\begin{equation*} \label{eq.master}
    \frac{d}{dt}\hat{\rho}=-\frac{i}{\hbar}[\hat{H},\hat{\rho}] 
+ \sum_j(\hat{L}_j\hat{\rho}\hat{L}_j^\dagger-\frac{1}{2}
\hat{L}_j^\dagger\hat{L}_j\hat{\rho}-\frac{1}{2}\hat{L}_j^\dagger\hat{L}_j),
\end{equation*}
the QSD equation is~\cite{P:book,SB:1997}:
\begin{equation} \label{eq.QSD} 
|\psi\rangle = -\frac{i}{\hbar}\hat{H}|\psi\rangle dt
+\sum_j(\langle\hat{L}_j^\dagger\rangle_\psi\hat{L}_j-
\frac{1}{2}\hat{L}_j^\dagger\hat{L}_j
-\frac{1}{2}\langle\hat{L}_j^\dagger\rangle_\psi
\langle\hat{L}_j\rangle_\psi)|\psi\rangle dt 
+\sum_j(\hat{L}_j-\langle\hat{L}_j^\dagger\rangle_\psi)
|\psi\rangle d\xi_j,
\end{equation}
and the quantum jump equation is:
\begin{equation} \label{eq.QJ} 
|\psi\rangle=-\frac{i}{\hbar}\hat{H}|\psi\rangle
dt+\sum_j(\frac{1}{2}\langle\hat{L}_j^\dagger\hat{L}_j
\rangle_\psi-\frac{1}{2}\hat{L}_j^\dagger\hat{L}_j)|\psi\rangle dt 
+\sum_j(\frac{\hat{L}_j}{\sqrt{\langle
\hat{L}_j^\dagger\hat{L}_j\rangle_\psi}}-1)|\psi\rangle dN_j.
\end{equation}
The QSD equation Eq.~\eqref{eq.QSD} is in the Ito form, which historically 
was called the nonlinear stochastic Langevin-Ito equation. Generally, for 
the Langevin equations of $N$ variables of the form
\begin{equation*} \label{eq:NL}
\dot{q_i}=h_i(\{q\},t)+g_{ij}(\{q\},t)\xi_j(t),
\end{equation*}
where $\{q\}=q_1,q_2,\dots,q_N$ and 
$\langle\xi_i(t)\rangle=0, \langle\xi_i(t)
\xi_j(t')\rangle=2\delta_{ij}\delta(t-t')$, the corresponding 
probability density function $W(\{x\},t)$ satisfies the Fokker-Planck
equation~\cite{R:book}
\begin{equation*} \label{eq:FK}
\frac{\partial W(\{x\},t)}{\partial t}=(-\frac{\partial}{\partial
    x_i}D_i(\{x\},t) +\frac{\partial^2}{\partial x_i \partial x_j}
D_{ij}(\{x\},t))W,
\end{equation*}
where the drift and diffusion coefficients are defined as
\begin{equation*} \label{eq:NFK}
\begin{split}
D_i(\{x\},t)&\equiv D_i^{(1)}(\{x\},t)=lim_{\tau\rightarrow0}\frac{1}{\tau}
\langle q_i(t+\tau)-x_i\rangle|_{q_k(t)=x_k}  
=h_i(\{x\},t)+g_{kj}(\{x\},t)\frac{\partial}{\partial x_k}g_{ij}(\{x\},t), \\
D_{ij}(\{x\},t)&\equiv D_i^{(2)}(\{x\},t)=\frac{1}{2}lim_{\tau\rightarrow0}\frac{1}{\tau}
\langle[q_i(t+\tau)-x_i][q_j(t+\tau)-x_j]
\rangle|_{q_k(t)=x_k} 
=g_{ik}(\{x\},t)g_{jk}(\{x\},t).
\end{split}
\end{equation*}
Note that $q_i(t+\tau)(\tau>0)$ is a solution of the Langevin equation, 
which has the sharp value $q_k(t)=x_k$ ($k=1,2\dots,N$) at time $t$. The
quantity $D_i^{(n)}(\{x\},t)=\frac{1}{n!}lim_{\tau\rightarrow0}\frac{1}{\tau}
\langle[q(t+\tau)-x]^n\rangle|_{q_k(t)=x_k}$ is the Kramers-Moyal expansion 
coefficients. For a process described by the Langevin equation with 
$\delta$-correlated Gaussian noise, all the Kramers-Moyal coefficients 
$D^{(n)}$ with $n\geq3$ vanish~\cite{R:book}. The physical significance 
is that the deterministic component of the Langevin equations contributes 
to the drift part in the evolution of the probability distribution while 
the stochastic component contributes to both the drift and diffusion 
evolution of the probability distribution.

In general, QSD represents a conditioned measurement experiment and the 
wave functions that it generates are normally localized about a point in 
the phase space. This fact can be exploited to improve the computational
efficiency~\cite{SBP:1995}. Say a wave function is localized about the
point $(q,p)$. We can represent it using the so-called \emph{excited 
coherent} basis states, $|q,p,n\rangle=D(q,p)|n\rangle$, instead of 
a large number of Fock states. Physically, this means that we exploit 
a moving basis that separates the wavefunction representation into a 
classical part $(q,p)$ and a quantum part $|q,p,n\rangle$, which is 
effectively a \emph{mixed} representation. The excited coherent states 
are defined through the coherent state displacement operator:
\begin{equation*}
    D(q,p)=exp\frac{i}{\hbar}(p\hat{Q}-q\hat{P}).
\end{equation*}
where $\hat{Q}$ and $\hat{P}$ are the position and momentum operators. 
The displacement operator can be defined using the creation/annihilation 
operator as
\begin{equation*}
    D(\alpha)=e^{\alpha^*\hat{a}^\dagger-\alpha\hat{a}},
\end{equation*} 
and the matrix element in Fock state is
\begin{equation*}
    \langle m|D(\alpha)|n\rangle=e^{\frac{1}{2}|\alpha|^2}\sqrt{\frac{m!}{n!}}
    (-\alpha^*)^{n-m}L_m^{n-m}(|\alpha|^2),
\end{equation*}
where $L_m^{n-m}(|\alpha|^2)$ is the associate Laguerre polynomials.

Suppose at $t=t_0$ the state of the system is localized about $(q_0,p_0)$, 
i.e.,
\begin{equation*}
    (q_0,p_0)=(\langle\psi(t_0)|\hat{Q}|\psi(t_0)\rangle,\langle\psi(t_0)
    |\hat{P}|\psi(t_0)\rangle).
\end{equation*}
After one time step, we have
\begin{equation*}
    (q_1,p_1)=(\langle\psi(t_0+\delta t)|\hat{Q}|\psi(t_0)+\delta 
        t\rangle,\langle\psi(t_0+\delta t)|\hat{P}|\psi(t_0
    +\delta t)\rangle)\ne(q_0,p_0)
\end{equation*}
We then shift the basis from $(q_0,p_0)$ to $(q_1,p_1)$, which can be
done through
\begin{equation*} \label{eq.delta}
    |\psi(t_0+\delta t)\rangle=D(-\delta q,-\delta p)|\psi(t_0)\rangle.
\end{equation*}
Besides the wavefunction, we need to transform the operators into the new
basis as well. The procedure is straightforward due to certain properties 
of the displacement operator:
\begin{equation*}
    \begin{split}
    D^\dagger(\alpha)\hat{a}D(\alpha)&=\hat{a}+\alpha \\
    D^\dagger(\alpha)\hat{a}^\dagger D(\alpha)&=\hat{a}^\dagger+\alpha^*
\end{split}
\end{equation*}
which changes the transformation of the Hamiltonian and the operators from 
two matrix multiplications to one matrix addition. In spite of the need 
to perform base transformation at each time step, the overall computational
speed is faster than that with the Fock state calculation.


\section*{Acknowledgements}

\noindent
We thank Dr.~L.~Huang and Mr.~ H.-Y.~Xu for helpful discussions.
This work was supported by AFOSR under Grant No.~FA9550-15-1-0151 and
by ONR under Grant No.~N00014-15-1-2405.

\section*{Author contributions}

\noindent
G.L.W., Y.C.L. and C.G. conceived and designed the research.
G.L.W. did the simulation. All participate in the result analysis. 
Y.C.L. wrote the paper with help from G.L.W. 

\section*{Completing financial interests:}
\noindent The authors declare no competing financial interests.

\end{document}